\documentstyle[referee]{l-aa}

\begin{document}

   \thesaurus{06         
              (03.11.1;  
               16.06.1;  
               19.06.1;  
               19.37.1;  
               19.53.1;  
               19.63.1)} 
   \title{Atomic data from the Iron Project}

   \subtitle{XLIII. Transition probabilities for Fe V}

   \author{Sultana N. Nahar\inst{1}, Franck Delahaye\inst{1},
Anil K. Pradhan\inst{1} \and C.J. Zeippen\inst{2}}
   \institute{ Department of Astronomy, The Ohio State University,
Columbus, OH 43210, USA \and
   UMR 8631 (associ\'ee au CNRS et \`a l'Universit\'e Paris 7) et DAEC,
 Observatoire de Paris, F-92195 Meudon, France}
   \offprints{S.N. Nahar}

\date{Received date; accepted date}

\def\etal{{\it et\thinspace al.}\ }
   \maketitle
\markboth{S.N. Nahar \etal: Transition probabilities for Fe V}{}

\begin{abstract}

 An extensive set of dipole-allowed, intercombination, and forbidden
transition probabilities for Fe~V is presented. The Breit-Pauli R-matrix
(BPRM) method is used to calculate $1.46 \times 10^6$ oscillator
strengths for the allowed and intercombination E1 transitions among
3,865 fine-structure levels dominated by configuration complexes with
$n \leq 10$ and $l \leq 9$. These data are complemented by an atomic
structure configuration interaction (CI) calculation using the
SUPERSTRUCTURE program for 362 relativistic quadrupole (E2) and magnetic
dipole (M1) transitions among 65 low-lying levels dominated by the
$3d^4$ and $3d^3 \ 4s$ configurations. Procedures have been developed
for the identification of the large number of fine-structure levels
and transitions obtained through the BPRM calculations. The target
ion Fe~VI is represented by an eigenfunction expansion of 19 fine-structure
levels of $3d^3$ and a set of correlation configurations. Fe~V bound
levels are obtained with angular and spin symmetries $SL\pi$ and
$J\pi$ of the (e~+~Fe~VI) system such that $2S+1$ = 5,3,1, $L \leq$ 10,
$J \leq 8$ of even and odd parities. The completeness of the calculated
dataset is verified in terms of all possible bound levels belonging to
relevant $LS$ terms and
transitions in correspondence with the $LS$ terms. The fine-structure
averaged relativistic values are compared with previous Opacity Project
$LS$ coupling data and other works. The 362 forbidden transition
probabilities considerably extend the available data for the
E2 and M1 transtions, and are in good agreement with those computed by
Garstang for the $3d^4$ transitions.

\end{abstract}

%

\section {Introduction}

 Astrophysical and laboratory applications often require large
datasets that are complete and accurate for comprehensive model
calculations of opacities (Seaton \etal 1994, the Opacity Project Team
1995, 1996), radiative forces (e.g. Seaton 1997, Hui-Bon-Hoa and Alecian 
1998, Seaton 1999), radiation transport in high-density
fusion plasmas, etc. 
The Opacity Project (OP) (The Opacity Project 1995, 1996; Seaton \etal
1994) produced large
datasets of transition probabilities for most astrophysically abundant
atomic systems in the close coupling approximation using the powerful
R-matrix method from atomic collision theory (Burke \etal 1971, Seaton
1987, Berrington \etal 1987). However, the calculations were carried out
in $LS$ coupling and the A-values were obtained neglecting relativistic
fine-structure. The LS multiplets may be divided into fine-structure
components using algebraic transformations. This has been done for a
number of atoms and ions using the OP data (or similar non-relativistic
calculations), including iron ions such as Fe~II (Nahar 1995), Fe~III
(Nahar and Pradhan 1996), and Fe~XIII (Nahar 1999). However, for 
such complex and heavy ions the neglect of relativistic effects may 
lead to a significant lack of precision, especially for weak transitions.

As an extension of the OP to include relativistic effects, the
present Iron Project (IP) (Hummer \etal 1993) employs relativistic
extensions of the R-matrix codes in the Breit-Pauli approximation
(Scott and Burke 1980, Scott and Taylor 1982, Berrington \etal
1995) to compute radiative and collisional atomic parameters.
Recently, several relativistic calculations of
transition probabilities have been carried out using the Breit-Pauli
R-matrix method (BPRM); e.g. for Fe~XXV and Fe~XXIV (Nahar and Pradhan
1999a), C~III (Berrington \etal 1998), Fe~XXIII ( Ram\'{i}rez \etal 
1999). These calculations produced  highly accurate oscillator strengths for
most transitions considered, within a few percent of experimental data
or other accurate theoretical calculations (where available).

However, in these relatively simple atomic systems the electron
correlation effects are weak and the configuration-interaction (CI,
in the atomic structure sense) is easier to account for than in the more
complex ions such as the low ionization stages of iron. In the present
report we present the results of a large-scale BPRM calculation for one
such ion, Fe~V, and discuss the accuracy and completeness of the
calculated data. An earlier work (Nahar and Pradhan 1999b) has decribed
certain important aspects of these calculations, in particular the
difficulty with the identification of levels and completeness of
fine-structure components within the LS multiplets.  The general aim of the
present work is two-fold: (i) to extend the IP work to the calculation of
relativistic transition probabilities for the complex low-z iron ions,
and (ii) to provide a detailed description of the extensive data tables
that should be essentially complete for most applications.

\section{Theory}

The theoretical scheme is described in earlier works (Hummer et al.
1993, Nahar and Pradhan, 1999a,b).  We sketch the basic points below.
In the coupled channel or close coupling (CC) approximation an atom
(ion) is described in terms of an (e + ion) complex that comprises of
a `target' ion, with N bound electrons, and a `free' electron that may
be either bound or continuum. The total energy of the system is either
negative or positive; negative eigenvalues of the (N + 1)-electron
Hamiltonian correspond to bound states of the (e + ion) system. In the
CC approximation the wavefunction expansion, $\Psi(E)$, for
a total spin and angular symmetry  $SL\pi$ or $J\pi$, of the (N+1)
electron system is represented in terms of the target ion states or
levels as:

\begin{equation}
\Psi_E(e+ion) = A \sum_{i} \chi_{i}(ion)\theta_{i} + \sum_{j} c_{j}
\Phi_{j},
\end{equation}

\noindent
where $\chi_{i}$ is the target ion wave function in a specific state
$S_iL_i\pi_i$ or level $J_i\pi_i$, and $\theta_{i}$ is the wave function
for the (N+1)th electron in a channel labeled as
$S_iL_i(J_i)\pi_i \ k_{i}^{2}\ell_i(SL\pi) \ [J\pi]$; $k_{i}^{2}$ is the
incident kinetic energy. In the second sum the $\Phi_j$'s are
correlation wavefunctions of the (N+1) electron system that (a)
compensate for the orthogonality conditions between the continuum and
the bound orbitals, and (b) represent additional short-range correlation
that is often of crucial importance in scattering and radiative CC
calculations for each $SL\pi$.

The BPRM method yields the solutions of the relativistic CC equations
using the Breit-Pauli Hamiltonian for the (N+1)-electron system to
obtain the total wavefunctions $\Psi_E(e+ion)$ (Hummer \etal 1993).
The BP Hamiltonian is
\begin{equation}
H_{N+1}^{\rm BP}=H_{N+1}+H_{N+1}^{\rm mass} + H_{N+1}^{\rm Dar}
+ H_{N+1}^{\rm so},
\end{equation}
where $H_{N+1}$ is the nonrelativistic Hamiltonian,
\begin{equation}
H_{N+1} = \sum_{i=1}\sp{N+1}\left\{-\nabla_i\sp 2 - \frac{2Z}{r_i}
        + \sum_{j>i}\sp{N+1} \frac{2}{r_{ij}}\right\},
\end{equation}
and the additional terms are the one-body terms, the mass correction, the
Darwin and the spin-orbit terms respectively. The spin-orbit interaction
splits the LS terms into fine-structure levels $J\pi$, where
$J$ is the total angular momentum. The positive and negative energy
states (Eq. 1) define continuum or bound (e~+~ion) states,

\begin{equation}
 \begin{array}{l} E = k^2 > 0  \longrightarrow
{\rm continuum~(scattering)~channel} \\  E = - \frac{z^2}{\nu^2} < 0
\longrightarrow {\rm bound~state}, \end{array}
\end{equation}
where $\nu$ is the effective quantum number relative to the core level.
Determination of the quantum defect ($\mu(\ell))$, defined as
$\nu_i = n - \mu(\ell)$ where $\nu_i$ is relative to the core level
$S_iL_i\pi_i$, is helpful in establishing the $\ell$-value associated
with a given channel level.

The $\Psi_E$ represents a CI-type wavefunction over a large number of
electronic configurations depending on the target levels included
in the eigenfunction expansion (Eq. 1). Transition matrix elements 
may be calculated with these wavefunctions, and the electron dipole 
(E1), electric quadrupole (E2), magnetic dipole (M1) or other operators
to obtain the corresponding transition probabilities. The present 
version of the BPRM codes implements the E1 operator to enable the 
calculation of dipole allowed and intercombination transition 
probabilities. The oscillator strength is proportional to the 
generalized line strength defined, in either length form or velocity 
form, by the equations
\begin{equation}
S_{\rm L}=
 \left|\left\langle{\mit\Psi}_f
 \vert\sum_{j=1}^{N+1} r_j\vert
 {\mit\Psi}_i\right\rangle\right|^2 \label{eq:SLe}
\end{equation}
and
\begin{equation}
S_{\rm V}=\omega^{-2}
 \left|\left\langle{\mit\Psi}_f
 \vert\sum_{j=1}^{N+1} \frac{\partial}{\partial r_j}\vert
 {\mit\Psi}_i\right\rangle\right|^2. \label{eq:SVe}
\end{equation}
In these equations $\omega$ is the incident photon energy in Rydberg 
units, and $\mit\Psi_i$ and $\mit\Psi_f$ are the bound wave
functions representing the initial and final states respectively.
The line strengths are energy independent quantities.

Using the energy difference, $E_{ji}$, between the initial and final
states, the oscillator strength, $f_{ij}$, for the transition can be
obtained from $S$ as

\begin{equation}
f_{ij} = {E_{ji}\over {3g_i}}S,
\end{equation}

\noindent
and the Einstein's A-coefficient, $A_{ji}$, as

\begin{equation}
A_{ji}(a.u.) = {1\over 2}\alpha^3{g_i\over g_j}E_{ji}^2f_{ij},
\end{equation}

\noindent
where $\alpha$ is the fine structure constant, and $g_i$, $g_j$ are
the statistical weight factors of the initial and final states,
respectively. In cgs units,
\begin{equation}
A_{ji}(s^{-1}) = {A_{ji}(a.u.)\over \tau_0},
\end{equation}

\noindent
where $\tau_0 = 2.4191 \times 10^{-17}$s is the atomic unit of time.

\section{Computations}

\subsection{The BPRM calculations}

The Fe~V wavefunctions are computed with eigenfunction expansions over
the `target' ion Fe~VI.
Present work employs a 19-level eigenfunction expansion of Fe VI 
corresponding to the 8-term $LS$ basis set of
$3d^3 (^4F$, $^4P$, $^2G$, $^2P$, $^2D2$, $^2H$, $^2F$, $^2D1)$, as
used in Nahar and Pradhan (1999b).
The target wavefunctions were obtained by Chen and Pradhan (1999) using
the Breit-Pauli version of the atomic structure code, SUPERSTRUCTURE 
(Eissner et al 1974). The bound channel set of
functions ${\Phi_j}$ in Eq. (1), representing additional (N+1)-electron 
correlation includes a number of Fe~V configurations, particularly from
the important n = 3 complex, i.e. $3s^23p^63d^4, 3p^63d^6, 3s^23p^53d^5,
3s^23p^43d^6$; the complete list of ${\Phi_j}$ for the n = 3 and 4
configurations is given in Chen and Pradhan (1999).

The Breit-Pauli calculations consider all possible fine-structure
bound levels of Fe V with (2$S$ + 1) = 1,3,5 and $L$ = 0 -- 10, 
$n\leq 10, \ \ell \leq n-1$, and $J \leq$ 8, and the transitions among
these levels. In the R-matrix computations, the calculated energies 
of the target levels were replaced by
the observed ones. The calculations are carried out using the BPRM codes 
(Berrington et al. 1995) extended from the Opacity Project codes 
(Berrington et al. 1987). 

STG1 of the BPRM codes computes the one- and two-electron radial
integrals using the one-electron target orbitals generated by
SUPERSTRUCTURE. The number of continuum R-matrix basis functions 
is chosen to be 12. The intermediate coupling calculations are 
carried out on recoupling these $LS$ symmetries in a 
pair-coupling representation in stage RECUPD. The computer 
memory requirement for this stage has been the
maximum as it carries out angular algebra of dipole matrix elements
of a large number of levels due to fine-structure
splitting. The (e + Fe~VI) Hamiltonian is diagonalized for each
resulting $J\pi$ in STGH.

\subsubsection{Energy levels and identification}

The negative eigenvalues of the (e + Fe VI) Hamiltonian correspond to the bound 
levels of Fe~V, determined using the code STGB.
Splitting of each target LS term into its fine-structure components also
increases the number of  Rydberg series of levels converging on to them.
These result in a large number of fine-structure levels in comparatively
narrow energy bands.  An order of magnitude finer mesh of effective
quantum number ($\Delta \nu$=0.001), compared to that needed for the
locating the bound $LS$ states, was needed to search for the
BP Hamitonian eigenvalues in order to avoid missing energy levels. 
The computational requirements
were, therefore, increased considerably for the intermediate coupling
calculations of bound levels over the LS coupling case by several 
orders of magnitude. The calculations take up to several CPU hours per
$J\pi$ in order to determine the corresponding eigenvalues in the
asymptotic program STGB.

The identification of the fine-structure bound levels computed in 
intermediate coupling using the collision theory BPRM method is rather 
involved, since they are labeled with quantum numbers related to
electron-ion scattering channels. The levels 
are associated with collision complexes of the (e + ion) system which, 
in turn, are initially identified only with their total angular momenta 
and parity, $J\pi$. A scheme has been developed (Nahar and Pradhan 
1999b) to identify the levels with complete spectroscopic information 
giving
\begin{equation}
C_t (\ S_t \ L_t)\ J_t~\pi_t n\ell \ [K] {\rm s}\ \ J \ \pi,
\end{equation}
and also to designate the levels with a possible $SL\pi$ symmetry
($C_t$ is the target configuration).

Most of the spectroscopic information of a computed level is extracted
from the few bound channels that dominate the wavefunction of 
that level. A new code PRCBPID has been developed to carry out the 
identification, including quantum
defect analysis and angular momentum algebra of the dominant channels.
Two additional problems are addressed in the identification work:
(A) correspondence of the computed fine-structure levels to the
standard $LS$ coupling designation, $SL\pi$, and (B)
completeness checks for the set of all fine-structure components
within all computed $LS$ multiplets. A correspondence between
the sets of $SL\pi$ and $J\pi$ of the same configuration are
established from the set of $SL\pi$ symmetries, formed from the
target term, $S_tL_t\pi_t$, $nl$ quantum numbers of the valence
electron, and $J\pi$ of the fine-structure level belonging to the $LS$
term. The identification procedure is described in detail in Nahar and 
Pradhan (1999b).

Considerable effort has been devoted to a precise and unique 
identification of levels.
However, a complex ion such as Fe~V involves many highly mixed levels
and it becomes difficult to assign a definite configuration and
parentage to all bound states.  Nonetheless, most of the levels have been
uniquely identified. In particular all calculated levels corresponding 
to the experimentally observed ones are correctly (and independently) 
assigned to their proper spectroscopic designation by the
identification procedure employed.

\subsubsection{E1 oscillator strengths}

The oscillator strengths and transition probabilities were obtained
using STGBB of the BPRM codes. STGBB computed the transition matrix
elements using the bound wavefunctions created by STGB, and the 
dipole operators computed by STGH. The
fine structure of the core and the (N+1) electron system
increased the computer memory and CPU time requirements considerably
over the LS coupling calculations.
About 31 MW of memory, and about one CPU hour on the Cray T94, was required
to compute the oscillator strengths for
transitions among the levels of a pair of $J\pi$ symmetries.
These are over an order of magnitude larger than those
needed for $f$-values in $LS$ coupling. The number of $f$-values
obtained from the BPRM calculations ranges from over 5,000, among 
$J$=8 levels, to over 123,000, among $J$=3 levels, for a pair of 
symmetries.

These computations required over 120 CPU hours
on the Cray T94. Total memory size needed was over 42 MW to diagonalise
the BP Hamilitonian. Largest computations involved a single $J\pi$ 
Hamiltonian of matrix size 3555, 120 channels, and 2010 configurations.

We have included extensive tables of all computed bound levels, and 
associated E1 A-values, with full spectroscopic identifications,
as standardized by the U.S. National Institute for Standards and 
Technology (NIST). In addition, rather elaborate (though rather
tedious) procedures are implemented to check and ensure completeness 
of fine-structure components within all computed LS multiplets. The 
complete data tables are available in electronic format. A sample of 
the datasets is described in the next section.

\subsection{SUPERSTRUCTURE calculations for the forbidden E2, M1
transitions}

The only available dataset by Garstang (1957) comprises of the E2,
M1 A-values for transitions within the ground $3d^4$ levels. The CI 
expansion for Fe~V consists of the configurations 
$(1s^22s^22p^63s^23p^6) \ 3d^4,3d^34s,3d^34p$ as the spectroscopic 
configurations and
$3d^34d,3d^35s,3d^35p,3d^35d,3d^24s^2,3d^24p^2,3d^24d^2,3d^24s4p,3d^24s4d$ as
correlation configurations. The eigenenergies
of levels dominated by the spectroscopic configurations are minimised
with scaling parameters $\lambda_{n\ell}$ in the  Thomas-Fermi-Dirac
potential used to calculate the one-electron orbitals in SUPERSTRUCTURE
(see Nussbaumer and Storey 1978):
$ { \lambda_{[1s-5d]}} \ = \ {1.42912,1.13633,1.08043,1.09387,1.07756,0.99000,
1.09616, 1.08171,}$

\noindent ${-0.5800,-0.6944,-1.0712,-3.0000} $.

There are 182 fine-structure levels dominated by the configurations
$3d^4$, $3d^34s$ and $3d^34p$, and the respective number of LS terms is
16, 32 and 80.  The $\lambda_{1s-3d}$ are minimised over the first
16 terms of $3d^4$, $\lambda_{4s}$ over 32
terms including $3d^34s$, and $\lambda_{4p}$, $\lambda_{5p}$ over all 80
terms. The $\lambda_{5s}$ and  $\lambda_{5d}$ are optimised over the
$3d^4$ terms to further improve the corresponding eigenfunctions.

 The numerical experimentation entailed a number of minimisation trials,
with the goal of optimisation over most levels. The final
set of calculated energies agree with experiment to within 10\%,
although more selective optimisation can lead to much better agreement
for many (but not all) levels. Finally, semi-empirical term energy 
corrections (TEC) (Zeippen \etal 1977) were applied to obtain the 
transition probabilities. This procedure has been successfully
applied in a large number of studies (see e.g. Bi\'emont \etal 1994).
The electric quadrupole (E2) and the magnetic dipole (M1)
transition probabilites, A$^q$ and A$^m$, are obtained using observed 
energies according to the expressions:

\begin{equation}
A_{j,i}^q(E2) = 2.6733 \times 10^3 (E_j - E_i)^5 {\cal S}^q(i,j)
sec^{-1},
\end{equation}

\noindent
and

\begin{equation}
A_{j,i}^m(M1) = 3.5644 \times 10^4 (E_j - E_i)^3 {\cal S}^m(i,j)
sec^{-1},
\end{equation}

\noindent
where $E_j > E_i$ (the energies are in Rydbergs), and ${\cal S}$ is 
the line strength for the corresponding transition.

\section {Results and discussion}

We have obtained nearly 1.5 $\times 10^{6}$ oscillator strengths for 
bound-bound transitions in Fe~V. To our knowledge there are no previous 
ab initio relativistic calculations for transition probabilities for 
Fe~V. The previous
Opacity Project data consists of approximately 30,000 LS
transitions. Therefore the new dataset of nearly 1.5 $\times 10^{6}$
oscillator strengths should significantly enhance the database, and
the range and precision of related applications, some of which we
discuss later.

We divide the discussion of energies and oscillator strengths in
the subsections below.

\subsection{Fine-structure levels from the BPRM calculations}

A total of 3,865 fine-structure bound levels of Fe V have been
obtained for $J\pi$ symmetries, 0 $\leq J \leq$ 8 even and odd 
parities. These belong to symmetries $2S+1$ = 5,3,1, 0 $\leq L \leq$ 
9, with $n \leq 10$ and 0 $\leq l \leq$ 9. The BPRM calculations 
initially yield only the energies and the total symmetry, $J\pi$, 
of the levels. Through an identification procedure based on the 
analysis of quantum defects and percentage channel contributions 
for each level in the region outside the R-matrix boundary 
(described in Nahar and Pradhan 1999b), the levels are assigned with 
possible designation of $C_t(S_tL_t)J_t\pi_tnlJ(SL)\pi$, which 
specifies the core or target configuration, $LS$ term and parity, 
and total angular momentum; the principal and orbital angular momenta, 
$nl$, of the outer or the valence electron; the total angular momentum, 
$J$, and the possible $LS$ term and parity, $SL\pi$, of the 
$(N+1)$-electron bound level. Table 1a presents a few partial sets 
of energy levels from the complete set available electronically.

\subsubsection{Computed order of levels according to $J\pi$}

Examples of fine structure energy levels are presented in sets of $J\pi$ in Table
1a where their assigned identifications are given. $N_J$ is the total 
number of energy levels for the symmetry $J\pi$ (e.g. there are 80 
levels with $J\pi$ = $0^e$, although the table presents only 25 of them). 
The effective quantum number, $\nu~=~z/\sqrt(E-E_t)$ where $E_t$ is the
energy of the target state, is also given for each level. 
The $\nu$ is not given for any equivalent electron level as it is undefined. 
An unidentifiable level is often assigned with a possible equivalent
electron level. In Table 1a, one level of $J\pi$ = $0^o$ is assigned to the
equivalent electron configuration, $3p^53d^5$. The assignment is based
on two factors: (a) the calculated $\nu$ of the level does not match with
that of any valence electon, and (b) the wavefunction is represented 
by a number of channels of similar percentage weights, i.e., no dominant
channel. The configuration $3p^53d^3$ corresponds to a large number 
of $LS$ terms. However, the level can not be identified with any 
particular term through quantum defect analysis. Hence it is 
designated as $^0S$, indicating an undetermined spectroscopic term.

\subsubsection{Energy order of levels}

In Table 1b a limited selection of energy levels is presented in a 
format different from
that in Table 1a. Here they are listed in ascending energy order
regardless of $J\pi$ values, and are grouped together within  the
same configuration to show the correspondence between the sets of
$J$-levels and the $LS$ terms. This format provides a check of
completeness of sets of energy levels in terms of $LS$ terms, and
also determines the missing levels. Levels grouped in such a manner also
show closely spaced energies, consistent with the fact that they are
fine-structure components with a given $LS$ term
designation. The title of each set in Table 1b lists all possible
$LS$ terms that can be formed from the core or target term, and outer
or the valence electron angular momentum. 'Nlv' is the total number
of $J$-levels that correspond to the set of $LS$ terms. The spin 
multiplicity ($2S+1$) and parity ($\pi$) are given next. The $J$ values for
each term is given with parentheses next to the corresponding $L$. At 
the end of the set of levels, 'Nlv(c)' is the total number of $J$-levels
obtained in the calculations. Hence, if Nlv = Nlv(c) for a set of levels
of the same configuration the set is designated as `complete'.

Most sets of fine-structure components between LS multiplets 
are found to be complete. High lying energy levels often belong to
incomplete sets. The possible $LS$ terms for each level is specified 
in the last column. It is seen that a level may possibly belong to several
$LS$ terms. In the absence any other criteriion, the proper term for the 
level may be assumed by
applying Hund's rule: with levels of the same spin multiplicity,
the highest $L$-level is usually the lowest. For example, of the two 
$J$=4 levels with terms $^3(F,G)$ in the second set, the first or 
the lower level could be $^3G$ while the second or the higher one 
could be $^3F$. It may be noted that this criterion is violated for a
number of cases in Fe~V due to strong CI. In Table 1b,
the upper sets of low energies are complete. The two lower sets are 
incomplete where a few levels are missing. The missing levels are 
also specified by the program PRCBPID.

\subsubsection{Comparison with observed energies}

Only a limited number of observed energy levels of Fe~V are available (Sugar
and Corliss 1985). All 179 observed levels were identified in a
straightforward manner by the program PRCBPID. The
present results are found to agree to about 1\% with the observed
energies for most of the levels (Table III, Nahar and Pradhan 1999b).
In Table 2, a
comparison is presented for the $3d^4$ levels. The
experimentally observed levels are also the lowest calculated levels
in Fe~V. The additional information in Table  2 is the level index,
$I_J$, next to the J-values. As the BPRM levels are designated with
$J\pi$ values only, the level index shows the energy position, in
ascending order, of the level in the $J\pi$ symmetry. It is necessary
to use the level indices to make the correspondence among the
calculated and the observed levels for later use.

Although the $LS$ term designation in general meets consistency checks,
it is possible that there is some uncertainty in the designations. The
spin multiplicities of the ion are obtained by the addition of
the angular momentum 1/2
of the outer electron to the total spin, $S_t$, of the target. The
higher multiplicity corresponds to the addition of +1/2, and the
lower one to the subtraction of -1/2. Typically the level with higher 
multiplicity
lies lower. Due to the large number of different channels representing the
levels of a $J\pi$ symmetry, it is
possible that this angular addition might have been interchanged for some
cases where the channels themselves are incorrectly identified.
Therefore, for example, a triplet could be represented by a singlet
and vice versa. This can affect the $L$
designation since a singlet can be assigned only to one single total $L$,
whereas a triplet can be assigned to a few possible $L$ values. For such
cases, the $LS$ multiplets may not represent the correct transitions.

We emphasize, however, that the present calculations are all in
intermediate coupling and the  LS coupling designations attempted in
this work are carried out only to complete the full spectroscopic
identification that may be of interest for (a) specialized users
such as experimentalists, and (b) as a record of all  possible
information (some of which may be uncertain) that can be derived from the
BPRM calculations for bound levels.

\subsection{E1 oscillators strengths from the BPRM Calculations}

The bound-bound transitions among the 3865 fine-structure levels of Fe~V
have resulted in $1.46\times 10^6$ oscillator strengths for dipole-allowed
and intercombination transitions.

\subsubsection{Calculated f-values for the allowed transitions}

Table 3 presents a partial set, in the format adopted, from the 
complete file of oscillator strengths. The two numbers at the 
beginning of the table are the nuclear charge (i.e. Z = 26) and the 
number of electrons ($N_{elc}$ = 22) for Fe~V. Below this line are 
the sets of transitions of a pair of symmetries $J\pi~-~J'\pi'$. 
The first line of each set contains values of $2J$, parity $\pi$ 
(=0 for even and =1 for odd), $2J'$ and $\pi'$. Hence in Table 3, 
the set of transitions given are among $J=0^e~-~J=1^o$.
The line following the transition symmetries specifies the number of
bound levels, $N_{Ji}$ and $N_{Jj}$, of the symmetries among which
the transitions occur. This line is followed by $N_{Ji}\times N_{Jj}$
transitions. The first two columns are the level indices, $I_i$ and
$I_j$ (as mentioned above) for the energy indices of the levels, and
the third and the fourth columns are their energies, $E_i$ and $E_j$,
in Rydberg units. The fifth and sixth columns are $gf_L$ and $gf_V$,
where $f_L$ and $f_V$ are the oscillator strengths in
length and velocity forms, and $g~=~2J+1$  ($J$ is the total
angular momentum of the lower level).
For the $gf$-values that are negative the lower level is $i$
(absorption) and for the
positive ones the lower level is $j$ (emission). The last column gives the
transition probability, $A_{ji}(sec^{-1})$. To obtain the identification
of the levels, Table 1a should be referrred to following $I_i$ and $I_j$.
For example, the second transition of Table 3 corresponds to
the intercombination transition $3d^4(^5D^e)(I_i=1) \rightarrow
3d^3(^4F^e)4p(^3D^o)(I_j=3)$.

\subsubsection{f-values with experimental energies}

As the observed energies are much more precisely known that the
calculated ones, the $f$ and $A$-values can be reprocessed with the
observed energies for some improvement in accuracy. Using the
energy independent BPRM line strength, $S$ (Eq. 7), the $f$-value
can be obtained as,
\begin{equation}
f_{ij} = S(i,j,BPRM){E_{ji}(obs)\over (3g_{i})}.
\end{equation}

Transitions among all observed levels have been so 
reprocessed. This recalculated subset consists of 3737 
dipole-allowed and intercombination transitions among the 179 
observed levels (a relatively small part of the present transition 
probabilities dataset). The calculated energy level indices 
corresponding to the observed levels for each $J\pi$ are 
listed in Table 4.  

A sample set of $f$- and $A$-values from the reprocessed transitions
are presented in Table 5a in $J\pi-J'\pi'$ order. Each transition is 
given with complete identification. The level index, $I_i$, for each energy
level is given next to the $J$-value for easy linkage to the energy and
$f$-files. In all calculations where large number of transitions are
used, the reprocessed $f$- and $A$-values should replace those in the
complete file (containing $1.46\times 10^6$ transitions).
For example, the $f$- and $A$-values for the first transition
$J=0^e(I_i=1)\rightarrow J=1^o(I_j=1)$ in Table 3 should be replaced
by those for the first transition in Table 5a. The overall replacement
of transitions can be carried out easily using the level energy index set
in Table 4.

\subsubsection{Spectroscopic designation and completeness}

The reprocessed transitions are further ordered in terms of their
configurations for a completeness check, and to obtain the $LS$ multiplet
designations.
A partial set is presented in Table 5b (the complete table is
available electronically). The completeness
depends on the observed set of fine-structure levels since transitions
have been reprocessed only for the observed levels. The $LS$ multiplets
are useful for various comparisons with existing values
where fine-structure transitions can not be resolved. 

 Semi-empirical atomic structure calculations have been carried out be
other workers (Fawcett (1989), Quinet and Hansen (1995)).
Present oscillator strengths are compared with available 
calculations by Fawcett (1989), the LS coupling R-matrix calculations from 
the OP (Butler, TOPbase 1993) and from the IP (Bautista 1996), 
for some low lying 
transitions. Comparison in Table 5b shows various degrees of 
agreement. Present $f$-values agree very well (within 10\%) with those 
by Fawcett for some fine-structure transitions while disagree 
considerably with the others within the same $LS$ multiplet. For example,
the agreement is good for most of the fine-structure transitions 
of $3d^4(^5D)\rightarrow 3d^3(^4F)4p(^5D^o)$, $3d^4(^5D)\rightarrow 
3d^3(^4P)4p(^5P^o)$, and $3d^42(^3P)\rightarrow 3d^3(^4F)4p(^3D^o)$ 
while the disagreement is large with other as well as with those of 
$3d^4(^5D)\rightarrow 3d^3(^4F)4p(^5F^o)$. 
The agreement of the present $LS$ multiplets
with the others is good for strong transitions such as $3d^4(^5D)\rightarrow
3d^3(^4F)4p(^5F^o,^5D^o,^5P^o)$, and $3d^42(^3P)\rightarrow
3d^3(^4F)4p(^3D^o)$, but is poor for the weak ones.

\subsubsection{Estimate of uncertainities}

The uncertainties of the BPRM transition probabilities for the allowed
transitons are expected to be within 10 \% percent for the
strong transitions, and 10-30\% for the weak ones. A measure of
the uncertainty can be obtained from the dispersion of the $f$-values
in length and velocity forms, which generally indicate deviations from
the `exact' wavefunctions (albeit with some exceptions).
Fig. 1 presents a plot of ${\rm log}_{10}{gf}$ values,
length vs. velocity, for the transitions $(J=2)^e-(J=3)^3$
of Fe~V.  Though most of the points lie close to the $gf_L = gf_V$ line, 
significant
dispersion is seen for $gf$ values smaller than 0.01. We should note
that the level of uncertainty may in fact be
less than the dispersion shown in Fig. 1; in the close coupling R-matrix
calculations the length formulation is likely to be more accurate than the
velocity formulaton since the wavefunctions are better represented in the
asymptotic region that dominates the contribution to the length form of
the oscillator strength.

\subsection{Forbidden transition probabilities}

The transition probabilities,  $A^q$ and $A^m$, for 362 forbidden 
E2 and M1 transitions are obtained using some semi-empirical 
corrections. The $A^q$ in general are much smaller than the $A^m$. 
Although $A^q$ is smaller, there are many cases where one 
or the other is negligible. Owing to the widespread use of the only 
other previous calculation by
Garstang (1957), it is important to establish the
general level of differences with the previous work. Table 6 gives a
detailed comparison. The agreement between the two sets of data is
generally good with a few noticeable discrepancies.

A partial set of the transition 
probabilities are given in Table 7 along with the observed wavelengths 
in microns. The full Table of forbidden transition probabilities is
available electronically.

\section {Conclusion}

To exemplify the future potential of computational spectroscopy with
the Breit-Pauli R-matrix method, complemented by related atomic structure
calculations, we present a fairly complete and large-scale set of
mainly {\it ab initio}
transition probabilities for a complex atomic system.
Level energies and fine-structure transition probabilities for Fe~V  are
presented
in a comprehensive manner with spectroscopic identifications.
We should expect these data to be particularly useful
for the calculation of monochromatic opacities and in the analysis of
spectra from astrophysical and laboratory sources where
non-local thermodynamic equilibrium (NLTE) atomic models with
many excited levels are needed.

 All data tables will be electronically available from the CDS archives, 
and via ftp from the first author at: nahar@astronomy.ohio-state.edu.

\begin{acknowledgements}
We would like to thank Dr. Werner Eissner for helpful comments and
general assistance with the BPRM codes.
This work was partially supported by
U.S. National Science Foundation (AST-9870089) and the NASA (NAG5 7903).
The computational work was
carried out on the Cray
T94 at the Ohio Supercomputer Center in Columbus, Ohio.
The collaboration between Columbus and Meudon benefitted from a visit to
the DAEC by AKP, with support from the Universit\'e Paris 7.

\end{acknowledgements}

\def\amp{{\it Adv. At. Molec. Phys.}\ }
\def\apj{{\it Astrophys. J.}\ }
\def\apjs{{\it Astrophys. J. Suppl. Ser.}\ }
\def\apjl{{\it Astrophys. J. (Letters)}\ }
\def\aj{{\it Astron. J.}\ }
\def\aa{{\it Astron. Astrophys.}\ }
\def\aasup{{\it Astron. Astrophys. Suppl.}\ }
\def\adndt{{\it At. Data Nucl. Data Tables}\ }
\def\cpc{{\it Comput. Phys. Commun.}\ }
\def\jqsrt{{\it J. Quant. Spectrosc. Radiat. Transfer}\ }
\def\jpb{{\it Journal Of Physics B}\ }
\def\pasp{{\it Pub. Astron. Soc. Pacific}\ }
\def\mn{{\it Mon. Not. R. astr. Soc.}\ }
\def\pra{{\it Physical Review A}\ }
\def\prl{{\it Physical Review Letters}\ }
\def\zpds{{\it Z. Phys. D Suppl.}\ }



\pagebreak

\begin{table}
\noindent{Table Ia. Identified fine-strucuture energy levels of Fe V.
$N_J$=total number of levels for the symmetry $J\pi$.
  \\ }
\scriptsize
\begin{tabular}{rllrrrcrl}
\noalign{\smallskip}
\hline
\noalign{\smallskip}
 i & $C_t$ & $S_tL_t\pi_t$ & $J_t$ & $nl$ & $J$ & $E_c(Ry)$ & $\nu$ & $SL\pi$
\\
 \noalign{\smallskip}
 \hline
 \noalign{\smallskip}
 \multicolumn{9}{c}{$N_J$= 80,~~~~~$J\pi$= $0^e$        } \\
 \noalign{\smallskip}
 \hline
 \noalign{\smallskip}
   1 & $3d4   $ &        &     &   & 0  &-5.56210E+00&      $^5 D e$ \\
   2 & $3d4 2 $ &        &     &   & 0  &-5.34705E+00&      $^3 P e$ \\
   3 & $3d4 2 $ &        &     &   & 0  &-5.16997E+00&      $^1 S e$ \\
   4 & $3d4 1 $ &        &     &   & 0  &-4.96162E+00&      $^3 P e$ \\
   5 & $3d4 1 $ &        &     &   & 0  &-4.43020E+00&      $^1 S e$ \\
   6 & $3d3   $ & $^4P^e$&1/2  &$4s$& 0 &-3.47060E+00& 2.62&$^3 P e$ \\
\hline
\end{tabular}
\end{table}


\pagebreak

\begin{table}
\noindent{Table 1b. Ordered and identified fine-structure energy
levels of Fe V. Nlv=total number of levels expected for the possible
$LS$ terms listed, and Nlv(c) = number of levels calculated. $SL\pi$
lists the possible $LS$ terms for each level (see text for details).
 \\ }
\scriptsize
\begin{tabular}{lcrrrcrl}
\noalign{\smallskip}
\hline
\noalign{\smallskip}
$C_t$ & $S_tL_t\pi_t$ & $J_t$ & $nl$ & $J$ & E(cal) & $\nu$ & $SL\pi$ \\
 \noalign{\smallskip}
 \hline
 \noalign{\smallskip}
\multicolumn{1}{l}{Nlv=~ 5,~ 5,e:} &
 \multicolumn{7}{l}{ F ( 5 4 3 2 1 )
             } \\
 \noalign{\smallskip}
 3d3      &(4Fe)& 3/2& 4s&  1& -3.73515E+00 &  2.59 &  5~ F e       \\
 3d3      &(4Fe)& 5/2& 4s&  2& -3.73238E+00 &  2.59 &  5~ F e       \\
 3d3      &(4Fe)& 5/2& 4s&  3& -3.72820E+00 &  2.59 &  5~ F e       \\
 3d3      &(4Fe)& 7/2& 4s&  4& -3.72275E+00 &  2.59 &  5~ F e       \\
 3d3      &(4Fe)& 9/2& 4s&  5& -3.71610E+00 &  2.59 &  5~ F e       \\
 \noalign{\smallskip}
 \multicolumn{8}{l}{Ncal=~ 5~ : set complete
                                                                       }
\\
\hline
\end{tabular}
\end{table}

\pagebreak

\begin{table}
\noindent{Table 2. Comparison of calculated BPRM energies, $E_c$, with
the observed ones (Sugar and Corliss 1985), $E_o$, of Fe V. $I_J$ is the
level index for the energy position in symmetry $J\pi$. \\
}
\scriptsize
\begin{tabular}{llrrrr}
\hline
\noalign{\smallskip}
\multicolumn{2}{c}{Level} & $J$ & $I_J$ & \multicolumn{1}{c}{$E_c$(Ry)} &
\multicolumn{1}{c}{$E_o$(Ry)} \\
\hline
            \noalign{\smallskip}
$3d^4 $        &     $^5D$ &    4 & 1  &       5.5493  &    5.5015 \\
$3d^4 $        &     $^5D$ &    3 & 1  &       5.5542  &    5.5058 \\
$3d^4 $        &     $^5D$ &    2 & 1  &       5.5580  &    5.5094 \\
$3d^4 $        &     $^5D$ &    1 & 1  &       5.5607  &    5.5119 \\
$3d^4 $        &     $^5D$ &    0 & 1  &       5.5621 &    5.5132 \\
$3d^42$        &     $^3P$ &    2 & 2  &       5.3247  &    5.2720 \\
$3d^42$        &     $^3P$ &    1 & 2  &       5.3389  &    5.2856 \\
\hline
\end{tabular}
\end{table}

\pagebreak

\begin{table}
\noindent{Table 3. Transition probabilities for Fe V (see text for
explanation).  \\ }
\scriptsize
\begin{tabular}{lrccccc}
\noalign{\smallskip}
\hline
\noalign{\smallskip}
\multicolumn{7}{l}{~~~26~~~~22} \\
\multicolumn{7}{l}{~~~~0~~~~0~~~~~~~~~2~~~~1} \\
80  & 236 & $E_i(Ry)$ & $E_j(Ry)$ & $gf_L$ & $gf_V$ & $A_{ji}(sec^{-1})$ \\
 \noalign{\smallskip}
   1&  1&-5.56210E+00&-3.14950E+00&-2.212E-01&-1.800E-01& 3.447E+09 \\
   1&  2&-5.56210E+00&-3.14082E+00&-5.660E-03&-4.178E-03& 8.884E+07 \\
   1&  3&-5.56210E+00&-3.13088E+00&-5.894E-02&-4.843E-02& 9.328E+08 \\
   1&  4&-5.56210E+00&-2.99367E+00&-8.675E-02&-7.425E-02& 1.532E+09 \\
   1&  5&-5.56210E+00&-2.97465E+00&-4.542E-03&-4.060E-03& 8.142E+07 \\
\hline
\end{tabular}
\end{table}

\begin{table}
\noindent{Table 4. Calculated energy level indices for various $J\pi$
symmetries; all allowed transitions have been reprocessed 
using observed energies. $n_j$ is the total number of $J\pi$
levels observed.
\\ }
\scriptsize
\begin{tabular}{ccrl}
\noalign{\smallskip}
\hline
\noalign{\smallskip}
$J$ & $\pi$ & $n_j$ & level indices \\
 \noalign{\smallskip}
\hline
 \noalign{\smallskip}
 0& e&    6 & 1,2,3,4,5,6 \\
 0& o&    6 & 1,2,3,4,6,7 \\
 1& e&   11 & 1,2,3,4,5,6,7,8,9,10,11 \\
 1& o&   19 & 1,2,3,4,5,6,7,8,9,10,11,12,13,14,15,16,17,18,19 \\
 2& e&   18 & 1,2,3,4,5,6,7,8,9,10,11,12,13,14,15,16,17,18 \\
 2& o&   24 & 1,2,3,4,5,6,7,8,9,10,11,12,13,14,15,17,18,19,20,21,22,23,24,25 \\
 3& e&   14 & 1,2,3,4,5,6,7,8,9,10,11,12,13,14 \\
 3& o&   24 & 1,2,3,4,5,6,7,8,9,10,11,12,13,14,15,16,17,18,19,20,21,22,23,24 \\
 4& e&   13 & 1,2,3,4,5,6,7,8,9,10,11,12,13 \\
 4& o&   18 & 1,2,3,4,5,6,7,8,9,10,11,12,13,14,15,16,17,18 \\
 5& e&    6 & 1,2,3,4,5,6 \\
 5& o&   11 & 1,2,3,4,5,6,7,8,9,10,11 \\
 6& e&    3 & 1,2,3 \\
 6& o&    5 & 1,2,3,4,5 \\
 7& o&    1 & 1 \\
\noalign{\smallskip}
\hline
\end{tabular}
\end{table}

\pagebreak

\begin{table}
\noindent{Table 5a. Sample set of reprocessed $f$- and $A$-values with
observed transition energies. $I_i$ and $I_j$ are the level indices.
\\ }
\scriptsize
\begin{tabular}{llllcrcrcccc}
\noalign{\smallskip}
\hline
\noalign{\smallskip}
\multicolumn{1}{c}{$C_i$} &\multicolumn{1}{c}{$C_j$} &
\multicolumn{1}{c}{$S_iL_i\pi_i$} & \multicolumn{1}{c}{$S_jL_j\pi_j$} &
$2J_i+1$ & $I_i$ & $2J_j+1$ & $I_j$ & $E_i(Ry)$ & $E_j(Ry)$
& $f$ & $A(sec^{-1})$ \\
 \noalign{\smallskip}
\hline
 \noalign{\smallskip}
 $3d^4         $ & $-~3d^3(4F)4p   $ & $^5D^e$ & $^5D^o$ &  1 &  1 &  3 &  2 &   5.5132 &   3.1540 &  5.515E-03 &  8.22E+07 \\
 $3d^4         $ & $-~3d^3(4F)4p   $ & $^5D^e$ & $^3D^o$ &  1 &  1 &  3 &  3 &   5.5132 &   3.1439 &  5.744E-02 &  8.63E+08 \\
 $3d^4         $ & $-~3d^3(4P)4p   $ & $^5D^e$ & $^5P^o$ &  1 &  1 &  3 &  4 &   5.5132 &   3.0195 &  8.420E-02 &  1.40E+09 \\
 $3d^4         $ & $-~3d^3(4P)4p   $ & $^5D^e$ & $^5D^o$ &  1 &  1 &  3 &  5 &   5.5132 &   3.0058 &  4.401E-03 &  7.41E+07 \\
 $3d^4         $ & $-~3d^3(4P)4p   $ & $^5D^e$ & $^3P^o$ &  1 &  1 &  3 &  6 &   5.5132 &   2.9911 &  8.365E-05 &  1.42E+06 \\
 $3d^4         $ & $-~3d^3(2P)4p   $ & $^5D^e$ & $^3P^o$ &  1 &  1 &  3 &  7 &   5.5132 &   2.9439 &  1.035E-03 &  1.83E+07 \\
 $3d^4         $ & $-~3d^3(2D2)4p  $ & $^5D^e$ & $^1P^o$ &  1 &  1 &  3 &  8 &   5.5132 &   2.9073 &  1.306E-08 &  2.38E+02 \\
\hline
\end{tabular}
\end{table}

\pagebreak

\begin{table}
\noindent{Table 5b. Fine-structure transitions, ordered in $LS$
multiplets, compared with previous values.
\\ }
\scriptsize
\begin{tabular}{llllrrrrll}
\noalign{\smallskip}
\hline
\noalign{\smallskip}
\multicolumn{1}{c}{$C_i$} &\multicolumn{1}{c}{$C_j$} & $S_iL_i\pi_i$ &
$S_jL_j\pi_j$ & $2J_i+1$ & $I_i$ & $2J_j+1$ & $I_j$ & $f_{ij}(present)$ &
$f_{ij}(others)$  \\
 \noalign{\smallskip}
\hline
 \noalign{\smallskip}
 3d4   &     -3d3(4F)4p & 5D0 &5F1 & 1 &1& 3 &1& 0.2154&0.163$^a$\\
 3d4   &     -3d3(4F)4p & 5D0 &5F1 & 3 &1& 3 &1& 3.790E-04&  \\
 3d4   &     -3d3(4F)4p & 5D0 &5F1 & 3 &1& 5 &3& 0.00136&  \\
 3d4   &     -3d3(4F)4p & 5D0 &5F1 & 5 &1& 3 &1& 0.04617&0.0126$^a$\\
 3d4   &     -3d3(4F)4p & 5D0 &5F1 & 5 &1& 5 &3& 0.05967&0.0596$^a$\\
 3d4   &     -3d3(4F)4p & 5D0 &5F1 & 5 &1& 7 &3& 0.01462&0.0138$^a$\\
 3d4   &     -3d3(4F)4p & 5D0 &5F1 & 7 &1& 5 &3& 0.006895&0.0274$^a$\\
\end{tabular}
\end{table}

\pagebreak

\begin{table*}
\noindent{Table 6. Comparison of $A$-values for forbidden transitions,
$A_{fi}^{cal}$, among $3d^4$ fine-structure levels with those,
$A_{fi}^G$, by Garstang (1957). \\ }
\begin{tabular}{llrrrrr}
\hline
\noalign{\smallskip}
 Transition & &  $\lambda_{vac}$ &  $E_{i}(cm^{-1})$
&   $E_{f}(cm^{-1})$  &  $A_{fi}^G(s^{-1})$ &
$A_{fi}^{cal}(s^{-1})$ \\
            \noalign{\smallskip}
            \hline
            \noalign{\smallskip}
$^5D_{1}$ & $^3P2_{0}$ & 4181.8 & 142.1 & 24055.4 & 1.30E+00 & 1.39E+00\\
$^5D_{0}$ & $^3P2_{1}$ & 4004.3 & 0.0 & 24972.9 & 1.30E-01 & 1.23E-01 \\
$^5D_{2}$ & $^3P2_{1}$ & 4072.4 & 417.3 & 24972.9 & 1.10E+00 & 1.07E+00 \\
$^5D_{1}$ & $^3P2_{2}$ & 3798.5 & 142.1 & 26468.3 & 3.60E-02 & 3.54E-02 \\
$^5D_{3}$ & $^3P2_{2}$ & 3896.3 & 803.1 & 26468.3 & 7.10E-01 & 7.08E-01 \\
$^5D_{4}$ & $^3H_{4}$ & 4228.4 & 1282.8 & 24932.5 & 1.10E-03 & 4.34E-03 \\
$^5D_{1}$ & $^3F2_{2}$ & 3756.8 & 142.1 & 26760.7 & 1.00E-01 & 1.04E-01 \\
\hline
\end{tabular}
\end{table*}

\begin{table*}
\noindent{Table 7. Forbidden transitions in Fe V. \\ \\}
\begin{tabular}{llllllrrrrr}
\hline
\noalign{\smallskip}
 $C_{i}$ & $C_{j}$ & $S_{i}L_{i}\Pi_{i}$ & $S_{j}L_{j}\Pi_{j}$ & $J_{i}$ & $J_{j}$  & $\lambda_{vac}$ & $E_{i} (cm^{-1})$ & $E_{j}(cm^{-1})$  &  $A_{ji}^{M1}(s^{-1})$ & $A_{ji}^{E2}(s^{-1})$ \\
            \noalign{\smallskip}
            \hline
            \noalign{\smallskip}
$3d^4$ & $3d^4$ & $^5D$ & $^5D$ & 0 & 1 & 703729.8 & 0.0 & 142.1 & 1.55E-04 & 0.00E+00\\
$3d^4$ & $3d^4$ & $^5D$ & $^5D$ & 0 & 2 & 239635.8 & 0.0 & 417.3 & 0.00E+00 & 1.17E-10\\
$3d^4$ & $3d^4$ & $^5D$ & $^5D$ & 1 & 2 & 363372.1 & 142.1 & 417.3 & 1.18E-03 & 1.10E-11\\
$3d^4$ & $3d^4$ & $^5D$ & $^5D$ & 1 & 3 & 151285.9 & 142.1 & 803.1 & 0.00E+00 & 1.00E-09\\
$3d^4$ & $3d^4$ & $^5D$ & $^5D$ & 2 & 3 & 259201.7 & 417.3 & 803.1 & 2.65E-03 & 1.29E-10\\
$3d^4$ & $3d^4$ & $^5D$ & $^5D$ & 2 & 4 & 115540.2 & 417.3 & 1282.8 & 0.00E+00 & 1.84E-09\\
$3d^4$ & $3d^4$ & $^5D$ & $^5D$ & 3 & 4 & 208463.6 & 803.1 & 1282.8 & 2.98E-03 & 4.19E-10\\
\hline
\end{tabular}
\end{table*}

\begin{table*}
\noindent{Table 7. Forbidden transitions in Fe V (Contd.) \\ \\}
\label{}
\begin{tabular}{llllllrrrrr}
\hline
\noalign{\smallskip}
 $C_{i}$ & $C_{j}$ & $S_{i}L_{i}\Pi_{i}$ & $S_{j}L_{j}\Pi_{j}$ & $J_{i}$ &
$J_{j}$  & $\lambda_{vac}$ & $E_{i} (cm^{-1})$ & $E_{j}(cm^{-1})$  &
$A_{ji}^{M1}(s^{-1})$ & $A_{ji}^{E2}(s^{-1})$ \\
            \noalign{\smallskip}
            \hline
            \noalign{\smallskip}

$3d^4$ & $3d^3(4F)$ & $^3P2$ & $^5F$ & 1 & 2 & 618.2 & 24972.9 & 186725.5 & 1.68E-07 & 1.63E-01\\
$3d^4$ & $3d^3(4F)$ & $^3P2$ & $^5F$ & 2 & 2 & 624.0 & 26468.3 & 186725.5 & 3.12E-07 & 6.09E-01\\
$3d^4$ & $3d^3(4F)$ & $^3P2$ & $^5F$ & 1 & 3 & 616.6 & 24972.9 & 187157.5 & 0.00E+00 & 4.94E-02\\
$3d^4$ & $3d^3(4F)$ & $^3P2$ & $^5F$ & 2 & 3 & 622.3 & 26468.3 & 187157.5 & 6.44E-07 & 3.56E-01\\
$3d^4$ & $3d^3(4F)$ & $^3P2$ & $^5F$ & 2 & 4 & 620.2 & 26468.3 & 187719.0 & 0.00E+00 & 5.32E-03\\
$3d^4$ & $3d^4$ & $^3H$ & $^3P2$ & 4 & 2 & 65112.6 & 24932.5 & 26468.3 & 0.00E+00 & 6.81E-10\\
$3d^4$ & $3d^4$ & $^3H$ & $^3H$ & 4 & 5 & 340831.6 & 24932.5 & 25225.9 & 6.60E-04 & 1.09E-13\\
\hline
\end{tabular}
\end{table*}

\begin{figure}
\caption{Comparison between the length and the velocity forms of
f-values for $(J=2)^e-(J=3)^3$ transitions in Fe V.}
\end{figure}

\end{document}